%% file: paper.tex
\documentclass[conference]{IEEEtran}
\usepackage[latin9]{inputenc}
\usepackage{amsmath,amssymb}
\usepackage{graphicx}
\usepackage{xcolor}

\ifCLASSINFOpdf
\else
\fi
\usepackage{array}

\usepackage{stfloats}
\begin{document}
%
\title{Adversarial Images through Stega Glasses}

\author{\IEEEauthorblockN{Benoît Bonnet }
\IEEEauthorblockA{Univ. Rennes, Inria, CNRS, IRISA \\
	Rennes, France \\
benoit.bonnet@inria.fr}
\and
\IEEEauthorblockN{Teddy Furon}
\IEEEauthorblockA{Univ. Rennes, Inria, CNRS, IRISA \\
	Rennes, France \\
	teddy.furon@inria.fr}
\and
\IEEEauthorblockN{Patrick Bas}
\IEEEauthorblockA{Univ. Lille, CNRS, Centrale Lille, UMR\\
	9189, CRIStAL, Lille, France \\
	patrick.bas@centralelille.fr}}


%


\maketitle

\begin{figure}[b]
\vspace{-0.3cm}
\parbox{\hsize}{\em
WIFS`2020, December, 6-9, 2020, New York, USA.
XXX-X-XXXX-XXXX-X/XX/\$XX.00 \ \copyright 2020 IEEE.
}\end{figure}

\begin{abstract}
This paper explores the connection between steganography and adversarial images.
On the one hand, steganalysis helps in detecting adversarial perturbations.
On the other hand, steganography helps in forging adversarial perturbations that are not only invisible to the human eye but also statistically undetectable.
This work explains how to use these information hiding tools for attacking or defending computer vision image classification.
We play this cat and mouse game with state-of-art classifiers, steganalyzers, and steganographic embedding schemes.
It turns out that steganography helps more the attacker than the defender.  
\end{abstract}


%
\IEEEpeerreviewmaketitle

\def \Xset{\mathcal{X}}
\def \x{\mathbf{x}}
\def \p{\mathbf{p}}
\def \q{\mathbf{q}}
\def \g{\mathbf{g}}
\newcommand{\fun}[1]{\mathsf{#1}}
\newcommand{\dfn}{:=}
\def \hc {\hat{c}}
\def \hp {\hat{p}}
\def \I {\mathbf{I}}
\def \ellb{\boldsymbol{\ell}}
\newcommand{\ted}[1]{{\color{red}{#1}}}
\newcommand{\pat}[1]{{\color{magenta}{#1}}}
\newcommand{\ben}[1]{{\color{blue}{#1}}}
\def\ie{\textit{i.e.}}
\def\Psuc{P_{\tiny{suc}}}

\input{Introduction}

\input{Sota}

\input{Embedding}

\input{Experimental}

\section{Conclusion}
This paper explores both sides of adversarial image detection with steganographic glasses. 

On the Defense side, we use SRNet~\cite{boroumand2018deep}, state-of-the-art in steganalysis to detect adversarial images.
Training it on images attacked with the basic FGSM shows impressive performance.
Detection also generalizes well even on the finest attacks such as PGD$_2$~\cite{Madry:2018aa} and CW~\cite{Carlini:2017ab}.

On the Attack side, our work on steganographic embedding is able to reduce dramatically the detection rates.
The steganographic embedding targets specific regions and pixels of an image to quantize the attack.
The distortion increases w.r.t. the original attack but remains imperceptible by the human eye (Fig.~\ref{fig:maxdisto}).
The main conclusion is that the field of steganography benefits more to the attacker than to the defender. 

Our future works will explore the effect of retraining detectors on adversarial images crafted with steganographic embedding towards an even more universal detector.






\bibliographystyle{IEEEtran}
\bibliography{AdvStega}
%

%
%
%
%
%

\end{document}

%% file: Introduction.tex

\section{Introduction}
\label{sec:Introduction}

Adversarial samples is an emerging field in Information Forensics and Security, addressing the vulnerabilities of Machine Learning algorithms. 
This paper casts this topic to the application of Computer Vision, and in particular, image classification.
A Deep Neural Network is trained to classify images depending on the type of object represented in the picture.
This is for instance the well-known ImageNet challenge encompassing a thousand of classes.
The state-of-the-art proposes impressive results as classifiers do a better job than humans with
less classification errors and much faster timings.
This may deserve the wording `Artificial Intelligence' 
as a computer now compete with humans on a difficult task.

The literature of adversarial samples reveals that these classifiers are vulnerable to specific image modifications.
For a given image, an attacker can craft a perturbation that triggers a wrong classification.
This perturbation is often a weak signal barely visible to the human eyes.
Almost surely, no human would incorrectly classify these adversarial images.
This topic is extremely interesting as it challenges the `Artificial Intelligence'  qualification too soon attributed to Deep Learning.

\subsection{Defenses}
We can find in the literature four types of defenses or counter-attacks to deal with adversarial contents:\\
\textbf{To detect:} Being barely visible does not mean that the perturbation is not statistically detectable.
This defense analyses the image and bypasses the classifier if detected as adversarial~\cite{Ma:2019aa}.\\
\textbf{To reform:} The perturbation looks like a random noise that may be filtered out.
This defense is usually a front-end projecting the image back to the manifold of natural images~\cite{Meng:2017aa}.\\
\textbf{To robustify:} At learning, adversarial images are included in the training set with their original class labels.
Adversarial re-training usually robustifies a `naturally' trained network~\cite{Madry:2018aa}.\\
\textbf{To randomize:} At testing, the classifier depends on a secret key or an alea.
This blocks pure white-box attacks~\cite{Taran:2019aa,Taran:2020aa}.

This paper deals with the first line of defense.
It is a pity that most papers proposing a defense do not seriously challenge it.
Security is often overclaimed as shown in~\cite{Athalye:2018aa,Carlini:2017aa}. 

\subsection{Connections with Information Hiding}
Paper~\cite{Quiring:2018qy} makes the connection between Adversarial Samples and Information Hiding (be it watermarking or steganography).
Both fields modify images (or any other type of media) in the spatial domain so that the content is moved to a targeted region of the feature space.
That region is the region associated to a secret message in Information Hiding or to a wrong class in Adversarial Sampling.
Indeed, paper~\cite{Quiring:2018qy} shows that Adversarial Sampling benefits from ideas proven efficient in Watermarking, and vice-versa.

This paper contributes to the same spirit by investigating what both Steganography and Steganalysis bring to Adversarial Sampling.
There are two natural ideas:\\
\textbf{Steganalysis} is the art of detecting weak perturbations in images.
This field is certainly useful for the defender.\\
\textbf{Steganography} is the art of modifying an image while being non-detectable.
This field is certainly useful for the attacker.\\

These two sides of the same coin allow to mount a defense and to challenge it in return.
This paper aims at revealing the status of the game between the attacker and the defender at the time of writing, \ie\ when both players use up-to-date tools:
state-of-the-art image classifiers with premium steganalyzers, and best-in-class steganography embedders.
As far as we know, this paper proposes three first time contributions:
\begin{itemize}
\item Assess the robustness of very recent image classifiers, EfficientNet~\cite{Tan:2019aa} and its robust version~\cite{Xie:2019aa}, 
\item Apply the best steganalyzer SRNET~\cite{boroumand2018deep} to detect adversarial images,
\item Use the best steganographic schemes to craft perturbations: 
HILL~\cite{li2014new} uses empirical costs, MiPod~\cite{sedighi2016content} models undetectability from a statistical point of view,  while GINA~\cite{li2015strategy,wang2019non} synchronizes embeddings on color channels. 
\end{itemize}

%% file: Sota.tex

\section{State of the art}
\label{sec:Sota}

\subsection{Steganalysis is Versatile}
Steganalysis has always been bounded to steganography, obviously.
Yet, a recent trend is to resort to this tool for other purposes than detecting whether an image conceals a secret message.
For instance, paper~\cite{Qiu:2014aa} claims the universality of SRM and LBP steganalyzers to detect image processing (like Gaussian blurring, gamma correction) and splicing. The authors of~\cite{Farooq:2017aa} used this approach during the IEEE IFS-TC image forensics challenge.
The same trend holds as well on audio forensics~\cite{Luo:2018aa}.
As for camera model identification, the inspiration from steganalysis (co-occurrences, color dependencies, conditional probabilities) is clearly apparent in~\cite{Tuama:2016aa}.

This reveals a certain versatility in steganalysis. It is not surprising since the main goal is to model and detect weak signals.
Modern steganalyzers are no longer based on hand-crafted features like SRM~\cite{fridrich2012rich}.
They are no more no less than Deep Neural Networks like XU-Net~\cite{xu2016structural} or SRNET~\cite{boroumand2018deep}.
The frontier between steganalysis and any two-class image classification problem (such as image manipulation detection) is blurred.
Yet, these networks have a specific structure able to focus on weak signal detection. They for example avoid pooling operations in order to preserve high frequency signals, they also need large databases combined with augmentation techniques and curriculum learning to converge~\cite{yousfi2019breaking}. 

This general-purpose based on steganalysis method  has some drawbacks.
It lacks fine-grained tampering localization, which is an issue in forensics~\cite{Fan:2015aa}.
Paper~\cite{Chen:2017aa} goes a step further in the cat-and-mouse game with an anti-forensic method: knowing that the defender uses a steganalyzer, the attacker modifies the perturbation (accounting for a median filtering or a contrast enhancement) to become less detectable.

As for adversarial images detection, this method is not new as well.
The authors of~\cite{Schottle:2018aa} wisely see steganalysis detection as a perfect companion to adversarial re-training.
This last mechanism fights well against small perturbation. It however struggles in correctly classifying coarser and more detectable attacks.
Unfortunately, this idea is supported with a proof of concept (as acknowledged by the authors): the steganalyzer is rudimentary, the dataset is composed of tiny images (MNIST). On the contrary, the authors of~\cite{Liu:2019aa} outline that steganalysis works better on larger images like ImageNet (ILSVRC-2016).
They however use a deprecated classifier (VGG-16) with outdated steganalyzers based on hand-crafted features (SPAM and SRM).

Conversely, adversarial samples recently became a source of inspiration for steganography: paper~\cite{Tang:2019aa} proposes the concept of steganography with an adversarial embedding fooling a DNN-based steganalyzer. 


\subsection{Adversarial Images}
This paper focuses on white-box attacks where the attacker knows all implementation details of the classifier.

To make things clearer, the classifier has the following structure: a pre-processing $\fun{T}$ maps an image $\I_o\in\{0,1,\ldots,255\}^n$ (with $n=3LC$, 3 color channels, $L$ lines and $C$ columns of pixels) to $\x_o = \fun{T}(\I_o)\in\Xset^n$, with $\Xset\dfn [0,1]$ (some networks also use $\Xset = [-1,1]$ or $[-3,3]$). This pre-processing is heuristic, sometimes it just divides the pixel value by $255$, sometimes this normalization is channel dependent based on some statistics (empirical mean and standard deviation).
This flattened vector $\x_o$ is fed the trained neural network to produce the estimated probabilities $(\hp_k(\x_o))_k$ of being from class $k\in\{1,\ldots,K\}$.
The predicted class is given by:
\begin{equation}
\hc(\x_o) = \arg \max_k \hp_k(\x_o).
\end{equation}
The classification is correct if $\hc(\x_o) = c(\x_o)$, the ground truth label of image $I_o$.

An \emph{untargeted} adversarial attack aims at finding the optimal point:
\begin{equation}
\label{eq:OptAdvAtt}
\x_a^\star = \arg \min_{\x:\hc(\x)\neq c(x_o)} \|\x - \x_o\|,
\end{equation}
where $\|\cdot\|$ is usually the Euclidean distance.

Discovering this optimal point is difficult because the space dimension $n$ is large.
In a white-box scenario, all attacks are sub-optimal iterative processes. They use the gradient of the network function efficiently computed thanks to the back-propagation mechanism to find a solution $\x_a$ close to $\x_a^\star$.
They are compared in terms of probability of success, average distortion, and complexity (number of gradient computations).
This paper considers well-known attacks: FGSM~\cite{Goodfellow:2015aa}, PGD (Euclidean version)~\cite{Madry:2018aa}, DDN~\cite{Rony:2019aa}, and CW~\cite{Carlini:2017ab} (ranked from low to high complexity).

As outlined in~\cite{Bonnet:2020aa}, definition~\eqref{eq:OptAdvAtt} is very common in literature, yet it is incorrect.
The goal of the attacker is to create an adversarial image $\I_a$ in the pixel domain.
Applying the inverse mapping $\fun{T}^{-1}$ is not solving the issue because this a priori makes non integer pixel values.
Rounding to the nearest integer, $\I_a = [\fun{T}^{-1}(\x_a)]$, is simple but not effective.
Some networks are so vulnerable (like ResNet-18) that $\fun{T}^{-1}(\x_a) - \I_o$ is a weak signal partially destroyed by rounding.
The impact is that, after rounding, $\I_a$ is no longer adversarial. 
Note that DDN is a rare example of a powerful attack natively offering quantized pixel values.

Paper~\cite{Bonnet:2020aa} proposes a post-processing $\fun{Q}$ on top of any attack that makes sure $\I_q = \fun{Q}(\fun{T}^{-1}(\x_a))$ is \textit{i)} an image (integral constraint), \textit{ii)} remains adversarial, and \textit{iii)} has a low Euclidean distortion $\|I_q - I_o\|$. This paper follows the same approach but adds another constraint: \textit{iv)} be non-detectable.

\subsection{Steganographic Embeddings}
Undetectability is usually tackled by the concept of costs in the steganographic literature: each pixel location $i$ of the cover image is assigned a set of costs $(w_i(\ell))_\ell$ that reflects the detectability of modifying the $i$-th pixel by $\ell$ quantum.
Usually, $w_{i}(0)=0$, $w_i(-\ell)=w_i(\ell)$, and $w_i(|\ell|)$ is increasing. The goal of the steganographer is to embed a message $\mathbf{m}$ while minimizing the empirical steganographic distortion:
\begin{equation}
\label{eq:StegoDist}
D(\ellb) \dfn \sum_{i=1}^{n} w_i(\ell_i).
\end{equation}
This is practically achieved using Syndrome Trellis Codes~\cite{filler2011minimizing}. Note that this distortion is additive, which is equivalent to consider that each modification yields to a detectability which is independent from the others. 

We propose to use the steganographic distortion (instead of $L_1$, $L_2$ or $L_\infty$ norms in adversarial literature) in order to decrease detectability.
There are strategies to take into account potential interactions between neighboring modifications. The image can first be decomposed into disjoint lattices to be sequentially embedded. And costs can then be sequentially updated after the embedding of every lattice
~\cite{li2015strategy}.  
This work uses three different families of steganographic costs.

The first one, HILL~\cite{li2014new}, is empirical and naive, but has nevertheless been widely used in steganography and is easy to implement. The cost map $\mathbf{w}$ associated to $\pm 1$ is computed using 2 low-pass averaging filters $\mathbf{L}_1$ et $\mathbf{L}_2$ of respective size $3\times 3$ et $15\times 15$ and one high pass filter $\mathbf{H}$:
\begin{equation}\label{cost:HILL}
\mathbf{w} = \frac{1}{|\mathbf{I}\ast\mathbf{H}|\ast \mathbf{L}_1}\ast \mathbf{L}_2,\text{with }
\mathbf{H}=\left[\begin{array}{rrr}
-1 & 2 & -1\\
2 & -4 & 2\\
-1 & 2 & -1
\end{array}\right].
\end{equation}


The second one, derived from MiPod~\cite{sedighi2016content}, assumes that the residual signal  is distributed as $\mathcal{N}(0,\sigma^2_i)$ for the original image,
and $\mathcal{N}(\ell_i,\sigma^2_i)$ for the stego image.
The variance $\sigma^2_i$ is estimated on each pixel using Wiener filtering and a least square approximation on a basis of cosine functions.
The cost is the log likelihood ratio between the two distributions evaluated at 0, i.e.:
\begin{equation}\label{cost:LRT}
w_i(\ell_i) = \ell_i^2/\sigma^2_i.
\end{equation}
Unlike the previous one, this model can handle modifications other than $\pm 1$.  

The last one is a cost updating strategy favoring coherent modifications between pixels within a spatial or color neighborhood.  It is called GINA~\cite{wang2019non} is derived from CMD~\cite{li2015strategy}. It splits the color images into 4 disjoint lattices per channel, i.e. 12 lattices. The embedding performs sequentially starting by the green channel lattices. The costs on one lattice is updated according to the modifications done on the previous ones as:
\def\sign{\mathsf{sign}}
\begin{equation}\label{cost:gina}
w_{i}'\left(\ell_i\right)=\frac{1}{9}w_{i}\left(\ell_i\right),\,\text{if }\,\sign(\ell_{i}) = \sign(\mu_i),
\end{equation}
with $\mu_{i}$ the average of the modifications already performed in the neighborhood of location $i$.


%% file: Embedding.tex

\section{Steganographic Post-Processing}
\label{sec:Embedding}
This section presents the use of steganography in our post-processing $\fun{Q}$ mounted on top of any adversarial attack. 

\subsection{About Steganography and Adversarial Examples}
Paper~\cite{Schottle:2018aa} stresses a fundamental difference:
Steganalysis has two classes, where the class `cover' distribution is given by Nature, whereas the class `stego' distribution is a consequence of designed embedding schemes. On the other hand, a \emph{perfect} adversarial example and an original image are distributed as by the class $\hc(\x_a)$ or $c(\x_o)$, which are both given by Nature.

We stress another major difference: Steganographic embedding is essentially a stochastic process.
Two stego-contents derived from the same cover are different almost surely. This is a mean to encompass the randomness of the messages to be embedded.
This is also the reason why steganographic embedders turns the costs $(w_i(\ell))_\ell$ into probabilities $(\pi_i(\ell))_\ell$ of modifying the $i$-th pixel by $\ell$ quantum. These probabilities are derived to minimize the detectability under the constraint of an embedding rate given by the source coding theorem:
\begin{equation}
R = - n^{-1}\sum_i \sum_{\ell_i} \pi_i(\ell_i) \log_2\left(\pi_i(\ell_i)\right)\,\mathrm{bits}.
\end{equation}


In contrast, an attack is a deterministic process always giving the same adversarial version of one original image.
 Adversarial imaging does not need these probabilities.

\subsection{Optimal post-processing}
Starting from an original image, we assume that an attack has produced $\x_a$ mapped back to $\I_a = \fun{T}^{-1}(\x_a)$.
The problem is that $\I_a\in[0,255]^n$, \ie\ its pixel values are a priori not quantized.
Our post-processing specifically deals with that matter, outputting $\I_q = \fun{Q}(\I_a)\in\{0,\ldots,255\}^n$.
We introduce $\p$ the perturbation after the attack and $\q$ the perturbation after our post-processing:
\begin{eqnarray}
\p &\dfn& \I_a - \I_o \in\mathbb{R}^n,\\
\ellb &\dfn& \I_q - \I_o \in\mathbb{Z}^n.
\end{eqnarray}
The design of $\fun{Q}$ amounts to find a good $\ellb$.
This is more complex than just rounding perturbation $\p$.

We first restrict the range of $\ellb$.
We define the degree of freedom $d$ as the number of possible values for each $\ell_i$, $1\leq i\leq n$.
This is an even integer greater or equal than 2.
The range of $\ell_i$ is centered around $p_i$.
For instance, when $d=2$, $\ell_i\in \{ \lfloor p_i\rfloor, \lceil p_i\rceil\}$.
In general, the range is given by
\def \pc{\lceil p_i\rceil}
\begin{equation}
\label{eq:Range}
\mathcal{L}_i\dfn \{ \pc - d/2,\ldots ,\pc - 1,\pc,\ldots,\pc+d/2-1\}.
\end{equation}
Over the whole image, there are $d^n$ possible sequences for $\ellb$.

We now define two quantities depending on $\ellb$.
The \emph{classifier loss} at $\I_q = \I_a -\p + \ellb$:
\begin{equation}
L(\ellb) \dfn  \log(\hp_{c_o}(\I_a -\p + \ellb)) - \log(\hp_{c_a}(\I_a -\p + \ellb)),
\end{equation}
where $c_o$ is the ground truth class of $\I_o$ and $c_a$ is the predicted class after the attack.
When the attack succeeds, it means that $\I_a$ is classified as $c_a\neq c_o$ because $\hp_{c_a}(\I_a) > \hp_{c_o}(\I_a)$ so that
$L(\p) < 0$. Our post-processing cares about maintaining this adversariality.
This constrains $\ellb$ s.t. $L(\ellb)<0$.

The second quantity is the \emph{detectability}.
We assume that a black-box algorithm gives the stego-costs $(w_i(\ell))_\ell$ for a given original image.
The overall detectability of $\I_q$ is gauged by $D(\ellb)$~\eqref{eq:StegoDist}.
In the end, the optimal post-processing $\fun{Q}$ minimizes detectability while maintaining adversariality:
\begin{equation}
\label{eq:ProblemInitial}
\ellb^\star = \arg \min_{\ellb:L(\ellb)<0} D(\ellb). 
\end{equation}

\subsection{Our proposal}
\label{sec:Proposal}
The complexity for finding the solution of~\eqref{eq:ProblemInitial}  a priori scales as $O(d^n)$.
Two ideas from the adversarial examples literature help reducing this.
First, the problem is stated as an Lagrangian formulation as in~\cite{Carlini:2017ab}:
\begin{equation}
\label{eq:LagrangianFormula}
\ellb_\lambda = \arg \min D(\ellb) + \lambda L(\ellb). 
\end{equation}
where $\lambda\geq0$ is the Lagrangian multiplier.
This means that we must solve this problem for any $\lambda$ and then find the smallest value of $\lambda$ s.t. $L(\ellb_\lambda)<0$.

Second, the classifier loss is linearized around $\I_a$, \ie\ for $\ellb$ around $\p$: $L(\ellb) \approx L(\p) + (\ellb-\p)^\top \g$, where $\g = \nabla L(\p)$.
This transforms problem~\eqref{eq:LagrangianFormula} into
\begin{equation}
\label{eq:ProbGivenLambda}
\ellb_\lambda = \arg \min \sum_{i=1}^n w_i(\ell_i) + \lambda (p_i-\ell_i).g_i. 
\end{equation}
The solution is now tractable because the functional is separable: we can solve the problem pixel-wise.
The algorithm stores in $d\times n$ matrix $W$ the costs, and in $d\times n$ matrix $G$ the values $((p_i - \ell_i).g_i)_i$ for $\ell_i\in\mathcal{L}_i$~\eqref{eq:Range}.
For a given $\lambda$, it computes $W+\lambda G$ and looks for the minimum of each column $1\leq i\leq n$.
In other words, it is as complex as $n$ minimum findings, each over $d$ values, which scales as $O(n\log d)$.

Note that for $\lambda=0$, $\fun{Q}$ quantizes $I_{a,i}$ `towards' $I_{o,i}$ to minimize detectability.
Indeed, if $\ell_i = 0$ is admissible ($0\in\mathcal{L}_i$ holds if $|p_i|  \leq d/2$),
then $\fun{Q}(I_{a,i}) = I_{o,i}$ at $\lambda=0$.

On top of solving~\eqref{eq:ProbGivenLambda}, a line search over $\lambda$ is required.
The linearization of the loss being a crude approximation, we make calls to the network to check that $\fun{Q}(\I_a)$ is adversarial:
When testing a given value of $\lambda$, $\ellb_\lambda$ is computed to produce $I_q$ that feeds the classifier.
If $I_q$ is adversarial then $L(\ellb_\lambda)<0$ and we test a lower value of $\lambda$ (giving more importance to the detectability), otherwise we increase it.
We use a binary search with a stopping criterion to control complexity of the post-processing. The search stops when two successive values of $\lambda$ are different by less than 1,000. Optimal $\lambda$ varies widely between different images. This criterion was empirically set to give both optimal value and short research time. 

\subsection{Simplification for quadratic stego-costs}
\label{sec:SimplQuadCost}
We now assume that the stego-costs obey to the following expression: $w_i(\ell) = \ell^2/\sigma_i^2$.
This makes the functional of~\eqref{eq:ProbGivenLambda} (restricted to the $i$-th pixel) equals to $\ell_i^2/\sigma_i^2 - \lambda g_i\ell_i + \lambda p_i$
which minimizer is $\tilde{\ell}_i = \lambda g_i\sigma_i^2 / 2$.

\begin{figure}[tbp]
\begin{center}
\includegraphics[width=0.7\columnwidth]{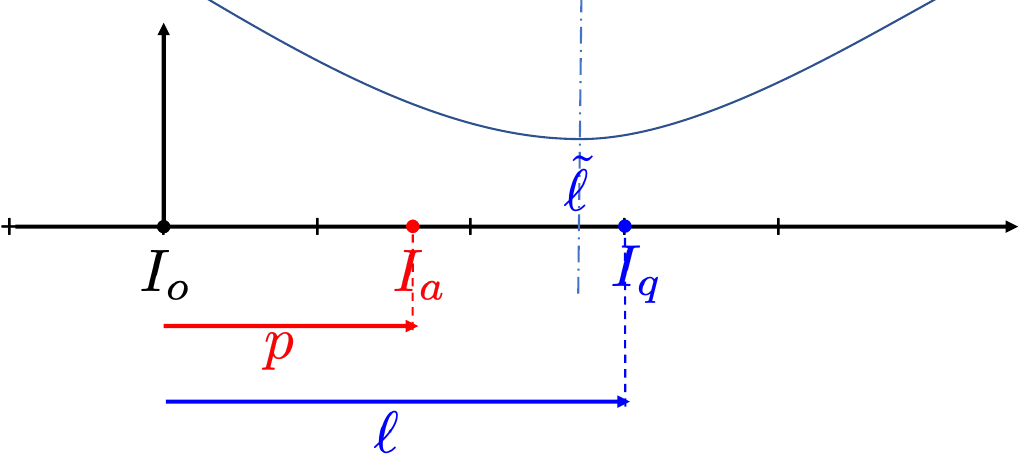}
\caption{Rounding the minimizer when the stego-cost is quadratic.}
\label{fig:minimum}
\end{center}
\end{figure}

Yet, this value a priori does not belong to $\mathcal{L}_i$~\eqref{eq:Range}.
This is easily solved because a quadratic function is symmetric around its minimum, therefore the minimum over $\mathcal{L}_i$ is its value closest to $\tilde{\ell}_i$ as shown in Fig.~\ref{fig:minimum}.
The range $\mathcal{L}_i$ being nothing more than a set of consecutive integers, we obtain a closed form expression:
\begin{equation}
\label{eq:CloseForm}
\ell_{\lambda,i} = \min(\max([\lambda g_i\sigma_i^2 / 2], \pc - d/2 ), \pc+d/2-1),
\end{equation}
where $[\cdot]$ is the rounding to the nearest integer. The post-processing has now a linear complexity.

In this equation, the min and max operate a clipping so that $\ell_{\lambda,i}$ belongs to $\mathcal{L}_i$.
This clipping is active if $\tilde{\ell}_i\notin\mathcal{L}_i$, which happens if $\lambda\geq \bar{\lambda}_i$ with
\begin{equation}
\bar{\lambda}_i :=
\begin{cases}
\left|\frac{2\lceil p_i\rceil -d}{g_i\sigma_i^2} \right|_+ &\text{if } g_i<0\\
\left|\frac{2\lceil p_i\rceil +d-2}{g_i\sigma_i^2} \right|_+&\text{if } g_i>0,
\end{cases}
\end{equation}
where $|a|_+ = a$ if $a>0$, $0$ otherwise.
This remark is important because it shows that for any $\lambda>\max_i \bar{\lambda}_i$, the solution $\ellb_\lambda$ of~\eqref{eq:CloseForm} remains the same due to clipping. Therefore, we can narrow down the line search of $\lambda$ to $[0, \max_i \bar{\lambda}_i]$.

%% file: Experimental.tex
\def \Acc{\mathrm{Acc}}
\def \TPR{\mathrm{TPR}}

\section{Experimental Investigation}
\label{sec:Experimental}

\subsection{Experimental setup}
Our experimental work uses 18,000 images from ImageNet of dimension 224$\times$224$\times$3. This subset is split in 1,000 for testing and comparing, 17,000 for training.
An image is attacked only if the classifier predicts its correct label beforehand.
This happens with probability equaling the accuracy of the network $\Acc$.
We measure $\overline{L_2}$ the average Euclidean distance of the perturbation $\ellb$ and $\Psuc$ the probability of a successful attack \emph{only over correctly labeled images}.

We attack the networks with 4 different attacks: FGSM~\cite{Goodfellow:2015aa}, PGD$_2$~\cite{Madry:2018aa}, CW~\cite{Carlini:2017ab} and DDN~\cite{Rony:2019aa}. All of the attacks are run in a \textit{best-effort} fashion with a complexity limited to 200 iterations. For FGSM and PGD$_2$ the distortion is gradually increased until the image is adversarial. For more complex CW and DDN, different sets of parameters are used on a total maximum of 200 iterations. The final attacked version is the adversarial image with the smaller distortion. 
DDN is the only attack that creates integer images.
The other 3 are post-processed either by the enhanced quantization~\cite{Bonnet:2020aa}, which is our baseline, or by our method explained in Sect.~\ref{sec:Proposal}.

The adversarial image detectors are evaluated by the true positive rate $\TPR$ when the false positive rate is fixed to $5\%$. 

\subsection{Robustness of recent classifiers: there is free lunch}
Our first experiment compares the robustness of the famous ResNet-50 network to the recent classifiers: the natural version of EfficientNet-b0~\cite{Tan:2019aa}  (Nat) and its robust version trained with AdvProp~\cite{Xie:2019aa} (Rob).
Note that the authors of~\cite{Xie:2019aa} apply adversarial re-training for improving accuracy.
As far as we known, the robustness of this version is not yet established.

Table~\ref{tab:RobustClassifier} confirms that modern classifiers are more accurate and more robust (lower $\Psuc$ and/or bigger $L_2$).
This is indeed a surprise: It pulls down the myth of `No Free Lunch' in adversarial machine learning literature~\cite{Tsipras:2018aa,Dohmatob:2019aa} (The price to pay for robustifying a network is pretendedly a lower accuracy).


\begin{table}[bt]
	\caption{Robustness of recent classifiers against PGD$_2$ followed by quantization~\cite{Bonnet:2020aa}}
		\label{tab:RobustClassifier}
	\begin{center}
		\begin{tabular}{ |l|c||c|c|} 
			\hline
			
			\\[-1em]
			& $\Acc$ (\%) & $\Psuc$ (\%)& $\overline{L_2}$   \\
			\hline 
			ResNet-50  & 80.0 & 97.2 &  81  \\ 
			`Nat' EfficientNet-b0~\cite{Tan:2019aa} & 82.8  &  88.0   & 88 \\ 
			`Rob' EfficientNet~\cite{Xie:2019aa}  & 84.3  & 71.8&  112  \\ 
			\hline
		\end{tabular}
	\end{center}
\end{table}	

\subsection{Detection with a Steganalyzer}

\begin{table}[!b]
	\caption{Detection of adversarial images with steganalyzers}
		\label{tab:DetectTruePosRate}
	\begin{center}
		\begin{tabular}{ |l|c|c||c|c|c|} 
			\hline
			
			\\[-1em]
			& $\Psuc$ & $\overline{L_2}$  & SRM(\%) &SCRMQ1(\%) & SRNet(\%) \\
			\hline 
			FGSM+\cite{Bonnet:2020aa}  & 89.7 & 286 &  72.00   & 83.3 & \bf{93.5}\\ 
			PGD$_2$+\cite{Bonnet:2020aa}  & 88.0 & 84 &  65.02   & 81.2 & \bf{93.3}\\ 
			CW+\cite{Bonnet:2020aa}  & 89.7  & 97&  68.78  & 83.6 & \bf{94.5}\\ 
			DDN & 83.2 & 186 &  79.53 & 91.9 & \bf{94.8}\\ 
			\hline
		\end{tabular}
	\end{center}
\end{table}

We use three steganalyzers to detect adversarial images.
Their training set is composed of 15,651 pairs of original and adversarial images.
The latter are crafted with \textit{best-effort} FGSM against natural EfficientNet-b0.

The first detector is trained on SRM feature vectors~\cite{fridrich2012rich}, with dimensions 34,671. SRM is a model that applies to only one channel. It is computed on the luminance of the image in our experimental work.
The classifier used to fit these high-dimensional vectors into two classes is the linear regularized classifier~\cite{cogranne2015ensemble}.
The second detector is based on the color version of SRM:  SCRMQ1~\cite{Goljan:2015aa} with dimension 18,157.
The classifier is the same.
The third detector is SRNet~\cite{boroumand2018deep}, one of the best detectors in steganalysis.
Training is performed on 180 epochs:
The first 100 with a learning rate of $10^{-3}$, the remaining 80 with $10^{-4}$.
Data augmentation is also performed during training.
First, there is a probability $p_1=0.5$ of mirroring the pair of images. Then, there is another probability $p_2=0.5$ of rotating them 90 degrees.

\textbf{The attacks}: Table~\ref{tab:DetectTruePosRate} shows that the probabilities of success $\Psuc$ are similar except for DDN (a larger complexity increases $\Psuc$ but it is not the aim of this study).
Note that PGD$_2$ and CW whose samples are quantized with~\cite{Bonnet:2020aa} are attacks as reliable as FGSM but with a third of the distortion.

\textbf{The detectors}:  Table~\ref{tab:DetectTruePosRate} gives also the $\TPR$ associated to the detectors. 
Although~\cite{Liu:2019aa} achieve good performances with SRM, we were not able to reproduce their results.
This could be due either to finer attacks or to the effect of quantization. 
Our results show that the detectors generalize well: although trained to detect images highly distorted by FGSM, they can detect as well and sometimes even better more subtle attacks like CW. Moreover, SRNet always outperforms SCRMQ1 and delivers an impressive accuracy. Table~\ref{tab:DetectTruePosRate} shows that PGD$_2$+\cite{Bonnet:2020aa} is the worst-case scenario for defense. The probability of fooling both the classifier EfficientNet-b0 and the detector SRNet combines to only 5.9\%.

\subsection{Post-processing with a Steganographic Embedder}
We now play the role of the attacker.
We use PGD$_2$ with best effort as the base attack to compare the detectability of four post-processings:
The non-steganographic insertion~\cite{Bonnet:2020aa} as a baseline,
HILL~\eqref{cost:HILL}, MiPod~\eqref{cost:LRT}, and GINA~\eqref{cost:gina}.
GINA uses the quadratic method explained in Sect.~\ref{sec:SimplQuadCost} sequentially over the 12 lattices.
Quadratic stego-costs are updated with CMD strategy~\eqref{cost:gina}.
Each lattice contributes to a 1/12 of the initial classification loss.

\begin{figure}[tbp]
	\begin{center}
		\begin{tabular}{cc} 
			
			\\[-0.8em]
			\includegraphics[height=1.5in,width=1.5in,]{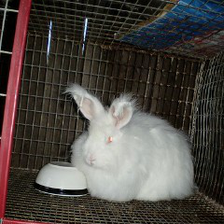} & \includegraphics[height=1.5in,width=1.5in,]{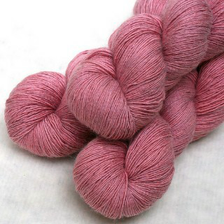} \\
			\lq Angora rabbit\rq & \lq woolen\rq \\
			\\[-0.8em]
			\includegraphics[height=1.5in,width=1.5in,]{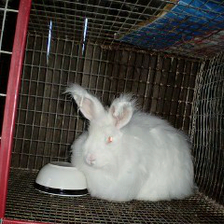} & \includegraphics[height=1.5in,width=1.5in,]{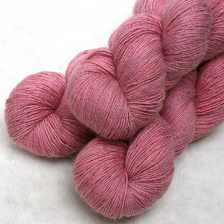}\\
			\lq hare\rq,  ${L_2}=488$  & \lq knot\rq,  ${L_2}=449$  \\ 

		\end{tabular}
		
	\end{center}
	
	\caption{Top row: Cover images with their label below. Bottom row: adversarial images with steganographic embedding GINA (d=4). Below them are their new label and the distortion}	
	\label{fig:maxdisto}
\end{figure}

Distortion increases with each method and along the degree of freedom $d$. Steganographic embedding therefore reduces detectability at the cost of increased distortion. From the attacker perspective, the best-case scenario with PGD$_2$ is with GINA at d=2 as seen on Table~\ref{tab:DetectTruePosRatesteg}. This scenario now has 69.9\% chance of fooling both the classifier and the detector on EfficientNet-b0. Fig.~\ref{fig:maxdisto} shows the two examples with highest distortion on EfficientNet-b0 that still fool SRNet. The added distortion remains imperceptible to the human eye even in these cases.

\begin{table}[bt]
	\caption{Undetectability of steganographic embedding\newline
	against the Naturel model (Nat) and its robust version (Rob).}
		\label{tab:DetectTruePosRatesteg}
	\begin{center}
		\begin{tabular}{ |l|c|c|c|c|c||c|c|c|c|} 
			\hline
			
			\\[-1em]
			 & $d$ &\multicolumn{2}{c|}{$\Psuc$ (\%)} &\multicolumn{2}{c||}{$\overline{L_2}$}  &\multicolumn{2}{c|}{SCRMQ1(\%)} &\multicolumn{2}{c|}{SRNet(\%)} \\
			 & & Nat & Rob &Nat & Rob &Nat & Rob &Nat & Rob\\
			 \hline
			\cite{Bonnet:2020aa} & 2& 88.0 & 71.8 & \bf{84} & \bf{112}& 81.2& 76.4 & 93.3 & 87.5\\ 
			HILL & 2& 88.0 &71.8 & 93 & 117 & 74.8& 66.3 & 86.1 & 77.6 \\ 
			HILL & 4& \bf{88.8} & \bf{72.6}& 105 & 129&  72.4& 72.4 & 85.5 &72.3\\ 
			MiPod & 2& 87.9 & 71.8 & 100  & 124 &  74.9& 64.3 & 84.0 & 76.1\\ 
			MiPod & 4& 88.2 & 72.2 & 114 & 137 &  72 & 57.0& 82.6 & 67.5\\ 
			GINA & 2 & 88.0 &71.8 & 168 & 181 &  5.4& \bf{3.0} & 44.2  & 33.5\\ 
			GINA & 4 & 88.2 & 71.9 & 232 & 243 &  \bf{3.8}&3.1 & \bf{20.7}  & \bf{14.2}\\ 
			\hline
		\end{tabular}
	\end{center}

\end{table}